\documentstyle[12pt]{article}
\newcommand{\Sbar}{\not{\!{\!S}}}
\newcommand{\pabar}{\not{\!\partial}}
\newcommand{\im}{\mbox{Im}}
\newcommand{\re}{\mbox{Re}}
\begin{document}

%
%
\input epsf 

\title{Resonant production of fermions in an axial background}
\author{Antonio L. Maroto$^1$ and Anupam Mazumdar$^{1,2}$\\${}$ \\
$^{(1)}$Astronomy Centre,
University of Sussex,\\
Falmer, Brighton BN1
9QJ, U.K.\\
$^{(2)}$Astrophysics Group, 
Imperial College,\\ 
Blackett Laboratory, Prince Consort Road,\\ 
London, SW7 2BZ, U.K}
\date{\today} 
\maketitle
\begin{abstract}
We consider the resonant production of fermions from an oscillating 
axial background. The classical evolution of the axial field is 
given by  that of a massive pseudovector field, as  suggested by
the renormalizability of the theory.  We     
look upon both the massive and the massless fermion production 
from a perturbative point of view. We obtain the corresponding spectrum and 
angular distributions for the different spins or helicities 
in the particular case of a spatial-like
axial field. We also extend our study to the non-perturbative regime
in the massless case 
and compare the results with the perturbative ones.
\end{abstract}




\section{Introduction}
The parametric resonance phenomenon plays a fundamental role in
the modern theories of reheating after inflation \cite{Linde}. 
In such theories,
after a period of inflation, driven generally by the potential
energy of the slowly rolling inflaton field, the inflaton
experiences a process of coherent oscillations around the minimum of the 
potential. Due to its coupling to the rest of the fields, those
oscillations can give rise to the production of exponentially 
large amount of bosons and subsequently to fermions through the decay 
of newly formed bosons, whose energy spectrum is characterized by  
resonance bands. Although, it is an efficient mechanism to convert the 
inflaton energy to the intermediate bosons, it is still far from complete 
and we still require to invoke 
the old idea of reheating where the bosons decay to various fermions 
and that subsequently leads to the thermalization of the universe. The 
standard picture of preheating has not been tested for  fields other than
inflatons, and therefore it is still unexplored  what would be
the effect of other fields that could be present in the early universe 
such as dilaton, axial or  moduli predicted by string theory.
It is a general belief that there would still be an amplification to one of
the dominant fields depending upon the values of coupling constants but it has
also been shown that for two scalar fields the system becomes chaotic
\cite{Maeda}. 

The production of particles from classical sources has other important
consequences apart from reheating. In fact, the same phenomenon
is responsible for the amplification of vacuum fluctuations in the
generation of gravitational waves \cite{graviton}. 
It has also been proposed as a mechanism
for the generation of primordial magnetic fields in the context of 
string cosmology \cite{Gio}. In this latter model, a dynamical dilaton
field plays the role of the external source. In addition,  
many other examples of
particle production
 based on alternative gravity theories can be found in the literature, 
such as
Brans-Dicke \cite{Ven}, higher dimensional models \cite{Garriga}, etc.

In this paper we do not intend to give a holistic picture of preheating;
rather we explore a different mechanism for the amplification
of vacuum fluctuations, it is based in the presence of an oscillating 
background axial field. Axial (or antisymmetric) fields appear naturally
in the bosonic sector of the low-energy string effective action
together with graviton and dilaton fields \cite{Witten,Tseytlin}. 
In fact, recently some
explicit solutions with non-trivial axion fields have been obtained
in the context of homogeneous, isotropic and
anisotropic models in string cosmology \cite{Copeland,shap}.
 These antisymmetric fields can be
interpreted as  torsion in pseudo-Riemannian geometry \cite{Tseytlin}
and, as we will show, this fact  provides a natural way of coupling them to 
other matter field.
This is also the case of supergravity theories 
which contain a non-symmetric part 
in the spin-connection that is determined by the gravitino field \cite{Neu}.

The presence of general chiral fields can give rise to a different 
kind of production from the usual scalar  fields. In particular,
the breaking of parity invariance could lead to a different production
of left and right fermions. In the present case, this will not happen
because our model is parity invariant. In addition, as far as rotational
invariance is broken by the  axial field, we will obtain an
anisotropic production of particles. In the usual picture of reheating
it is expected that such initial anisotropy is rapidly erased by 
subsequent decays of the fermions produced in arbitrary directions.
On the other hand, as far as the background field could have a non-vanishing
angular momentum, we should
also pay attention to the spin distribution of the produced particles.
Again in our case, we will consider a very simple homogeneous background
with vanishing total angular momentum and we will not generate
any net spin.

The paper is organized as follows: in Section 2, we study  
the classical equations of motion
for the axial field and its coupling to fermions. Section 3 is devoted to 
 the 
perturbative calculation valid in the limit of small axial field for both
massive and massless fermions.
In Section 4 we extend our calculation to the non-perturbative 
regime in the masless limit and compare our results with the previous ones.
In Section 5 we include the main conclusions of the work.

\section{Minimal coupling and  the axial field dynamics}

The torsion field can be seen to be minimally coupled
to fermions by means of the Einstein Equivalence Principle \cite{Hehl,DoMa}. 
Thus, the
Dirac lagrangian for a single fermion field wih 
mass $m$ in the presence of curvature and torsion is 
given by:
\begin{eqnarray}
{\cal L}=\overline\psi \left(i\gamma^\mu\left(\partial_\mu
+\Omega_\mu+igS_\mu 
\gamma_5\right)-m\right)\psi 
\label{lag}
\end{eqnarray}
where  $S_{\rho}=\epsilon_{\mu\nu\lambda\rho}T^{\mu\nu\lambda}$ is the
torsion pseudotrace, $\Omega_\mu$ the spin-connection and the
coupling constant is fixed $g=1/8$, (we will work with a general $g$
in order to extend our results to other models in which $S_\mu$
is not interpreted as torsion). 
 Despite the
presence of the $\gamma_5$, the lagrangian is parity invariant
since $S_\mu$ is a pseudovector. 
As a consequence of this coupling, the axial field  
can act as a source
of fermion creation. Notice that this coupling
is dictated by the Einstein Equivalence Principle, unlike the
usual inflaton couplings which are introduced ad hoc.
The problem of fermion production has received attention only
very recently in the  works of Greene and Kofman \cite{GK} and 
Baacke et al. \cite{Ba}. 
They have
studied for the first time the problem of reheating of fermions 
from the coherent oscillations of the inflaton field. They showed
how the limit on occupation number imposed by Pauli blocking
was saturated by the resonant production. 

The first problem we have to deal with regarding the creation of particles
is that of the classical dynamics of the 
torsion field. There are several models proposed in the literature
for which torsion appears in the classical action only as a mass term, thus
for instance:
the Einstein-Cartan theory  considers the Einstein-Hilbert
action but replaces the scalar curvature built out of the Levi-Civita
connection by the scalar curvature built out of an arbitrary
connection with torsion. In this theory torsion is a non-propagating
field that appears in the lagrangian simply in a mass term
of the form $M_P^2 S^2$, with $M_P$ the Planck mass.
This is also the case of the bosonic sector of the low-energy
string effective action.   
However, if we consider the coupling of torsion to 
quantum fermionic fields in (\ref{lag}),
the conditions of unitarity and renormalizability
automatically requires the existence of a kinetic term for the axial 
field, together
with the previous mass term, i.e, 
they impose
the axial field to behave like an abelian massive axial-vector
\cite{shapiro,DoMapre}. In fact, 
the vacuum divergences generated by the above lagrangian in (\ref{lag})
are given  by:
\begin{eqnarray}
S_{div}[S]=\frac{\Delta}{(4\pi)^2}\int d^4x\left(
-\frac{1}{192}S_{\mu\nu}S^{\mu\nu}+\frac{m^2}{16}S_\mu S^\mu\right)
\label{div}
\end{eqnarray}
where $S_{\mu\nu}=\partial_\mu S_\nu-\partial_\nu S_\mu$,  
$\Delta=N_\epsilon+\log (\mu^2/m^2)$ with the poles parametrized as
usual in dimensional regularization by
$N_\epsilon=2/\epsilon+\log 4\pi-\gamma$ and
$\mu$ is the renormalization scale. Notice that there is no
 $S^4$ term although it has the same dimension.
When instead of a single
fermion we have more than one, the previous action is replaced
by the sum of the actions corresponding to each single
fermion. Therefore by an appropriate 
renormalization
of mass and wave function we can take as the classical lagrangian
for torsion that of the abelian massive gauge field.
This is the minimal lagrangian for torsion that ensures the 
renormalizability of the fermionic sector.  This model is
analogous to the inflaton model with a quadratic potential, but
replacing the scalar field by a pseudo-vector.
Notice that unlike the vector abelian case, the presence of
massive fermions is incompatible with a gauge invariance 
associated to the axial field.
The corresponding
equation of motion is nothing but the Proca equation, i.e:
\begin{eqnarray}
\partial_\mu S^{\mu\nu}+m_s^2S^\nu=0
\label{proca}
\end{eqnarray}
with $m_s$ the mass of torsion. It is now easy to find solutions
for this equation. In particular, when torsion is spatial-like and
only depends on time we have the following 
oscillatory solutions: $S^i(t)=\hat S^i \sin(m_s (t-t_0))$
with $\hat S^i$ constants. As is well-known,
 this kind of periodic functions
 can give rise to the parametric resonance pheonomenon.

As in the inflaton models, the production of particles induces the 
damping in the torsion
oscillations. This can be described either by including a friction term in
the equations of motion or by means of the vacuum polarization 
correction to the torsion mass. The vacuum polarization correction
has been calculated in a previous paper \cite{axial} and from its 
imaginary part
we get that:
\begin{eqnarray}
\Gamma\equiv\frac{\im \Pi(m_s)}{m_s}\propto\frac{g^2m^2}{m_s}
\end{eqnarray}
where $m^2$ is the mass of the fermion squared or in the
case of several fermions present, it is given by $m^2=\sum_i m_i^2$.
Accordingly the classical evolution can be taken as:
\begin{eqnarray}
S^i(t)=\hat S^i\sin(m_s(t-t_0))e^{-\frac{1}{2}\Gamma (t-t_0)}
\label{evol}
\end{eqnarray}
For the sake of simplicity we do not consider the expansion of the universe
and accordingly all the calculations will be done in Minkowski
space-time. We will also simplify further and take $S^i$ pointing
in the z-direction.
As usual in particle production calculations, we will impose our
initial conditions on the background fields in such a way that it is
possible to define asymptotic $in$ and $out$ vacuum states. With that purpose
we will assume in next section that when $t\rightarrow -\infty$ 
the axial field is
zero, i.e we will take $S^i=0$ for $t<t_0$ and $S^3(t)$ given 
by (\ref{evol}) for $t\geq t_0$, as we will show the
result will not depend on the particular value of $t_0$ but only
on the amplitiude of the oscillations $\hat S^i$ and decay rate $\Gamma$.

\section{Perturbative approach}

{\bf Massive fermions}

Let us consider the
Dirac equation in the presence of the axial field:
\begin{eqnarray}
\label{Dirac1}
\left( i \gamma^{\mu }\partial_{\mu } -gS_{\mu }\gamma^{\mu }
\gamma_{5} -m \right)\Psi =0
\label{direq}
\end{eqnarray}
It is to be noted that the form of 
the coupling does not allow us to 
disentangle the different spin modes, unlike in the case of
Greene and Kofman.
In addition, it  cannot be reduced in an easy way to the known 
form of Mathieu equation or Lame equation,
which has the known  stability and the instability bands. During instability
bands the occupation number grows exponentially fast in the case of bosons.
This keep us from finding the exact solution of the equation. However,
it is possible to perform a perturbative analysis which provides
part of the resonant behaviour of the equation.
 There exists various approximation schemes to apply
for the bosons as well as the fermions; we will follow
 that in \cite{BiDa}.
Multiplying (\ref{Dirac1}) with $\left( i \gamma^{\mu }\partial_{\mu } 
+gS_{\mu }\gamma^{\mu } \gamma_{5} +m \right)$, we get
\begin{eqnarray}
\label{Dirac2}
\left[\partial_{\mu }\partial^{\mu } +ig \gamma^{\mu}
\gamma^{\nu }\gamma_{5} \partial_{\mu } S_{\nu }  +ig  S_{\nu }
\{ \gamma^{\mu }, \gamma^{\nu } \} \gamma_{5} \partial_{\mu } 
 -g^2S^{2} + 2gm \Sbar \gamma_5 +m^2\right]  
 \Psi =0 
\end{eqnarray}
Since there is no dependence on position in the Dirac
operator, we can write the general solution $\Psi$ as:
\begin{equation}
\label{Sol1}
\Psi(t,\vec x) =  \psi_{\vec k,s}(t) e^{i\vec{k}\vec{x}} 
\end{equation}
where $s$ is a spin index. 
As we mentioned before we will take  $S_\mu \equiv (0, 0, 0, S_{3})$.
Eq.(\ref{Dirac2}) reduces to:
\begin{eqnarray}
\left(\partial_{0}^{2} +\vec{k}^{2} +ig\dot{S_{3}}\gamma^{0}\gamma^{3}
\gamma_{5} -2g k_{3}S_{3}\gamma_{5} +m^{2} + g^2S_{3}^{2}\,
 +2gmS_{3}\gamma^{3}\gamma_{5} \right)\psi_{k,s}(t) = 0
\end{eqnarray}
We introduce the plane-wave solutions for massive spinors with
positive and negative frequency, and with a definite spin
projection along the z-direction:
\begin{eqnarray}
U_{\vec k,s}(t,\vec{x}) &=& 
\frac{1}{\sqrt{2\omega}}u(\vec k,s)e^{i\vec{k}\vec{x} 
-i\omega t}\, \\
V_{\vec k,s}(t,\vec{x}) &=& 
\frac{1}{\sqrt{2\omega}}v(\vec k,s)e^{-i\vec{k}\vec{x}
+i\omega t}
\label{planew}
\end{eqnarray}
where $s=\pm$ and  $\omega^2=\vec k^2+m^2$, 
$u(\vec{k},s)$ and $v(\vec{k},s)$ satisfy the following normalization 
conditions:
\begin{eqnarray}
u^{\dagger}(\vec{k},r)u(\vec{k},s) &=& 2\omega \delta_{r s}\,\\
v^{\dagger}(\vec{k},r)v(\vec{k},s) &=& 2\omega \delta_{r s}\, \\
u^{\dagger}(-\vec{k},r)v(\vec{k},s)&=&0\, \\
v^{\dagger}(-\vec{k},r)u(\vec{k},s)&=&0
\end{eqnarray}
where $u(\vec{k},s)$ and $ v(\vec{k},s)$ are defined as:
\begin{eqnarray}
u(\vec{k},s)=\sqrt{\omega+m}\left(\begin{array}{c}
\chi_s\\
\frac{\vec{\sigma}.\vec{k}}{\omega+m}\chi_s  \
\end{array}\right), \;\;
v(\vec{k},s)=\sqrt{\omega+m}\left(\begin{array}{c}
\frac{\vec{\sigma}.\vec{k}}{\omega+m}\chi_s\\
\chi_s\
\end{array}\right)
\end{eqnarray}
and the Dirac spinors have the simple form:
\begin{eqnarray}
\chi_{+} =\left( \begin{array}{c}
1\\
0\
\end{array}\right), \;\;
\chi_{-} =\left( \begin{array}{c}
0\\
1\
\end{array}\right)
\end{eqnarray}
We will work in this section in the Dirac representation for which 
the $\gamma$ matrices appear as:
\begin{eqnarray}
\gamma^{0}=\left( \begin{array}{cc}
1& 0\\
0& -1\
\end{array}\right), \;\;
\gamma^{i}=\left( \begin{array}{cc}
0& \sigma^{i}\\
-\sigma^{i}& 0\
\end{array}\right), \;\; 
\gamma_{5}=\left( \begin{array}{cc}
0& -1\\
-1& 0\
\end{array}\right)
\end{eqnarray}
Let us consider a given solution of the Dirac equation $\psi^{in}$
that initially,
i.e. $t\rightarrow -\infty,$ behaves like a plane wave with positive
energy and spin s. In the remote future, where the interaction
is switched off, such a solution will evolve into
a linear combination of positive and negative frequences with a certain
probability for spin flip, that is:
\begin{eqnarray}
\psi^{in}_{\vec{k} s}(t,\vec{x})\stackrel{t\rightarrow \infty}{\rightarrow}
\sum_{s\prime} \alpha_{\vec{k}ss\prime}
U_{\vec{k}s\prime} + \beta_{\vec{k}ss\prime}V_{-\vec{k}s\prime} 
\label{bog}
\end{eqnarray}
The Bugolubov coefficients for  fermions satisfy the following relation:
\begin{eqnarray}
\sum_{s\prime }(|\alpha_{\vec{k}ss^\prime}|^{2} 
+ |\beta_{\vec{k}ss^\prime}|^{2})=1
\end{eqnarray}
It is possible to transform the differential equation
into an integral equation that allows a perturbative
expansion:
\begin{equation}
\psi_{\vec{k} s}^{in}(t,\vec{x})= U_{\vec{k} s} +\frac{1}{\omega}
\int_{-\infty}^{t}M_{\vec{k}}\sin(\omega(t-t^{\prime}))\psi_{\vec{k}s}^{in}
(t',\vec{x})dt^{\prime}
\label{pert}
\end{equation}
where the interaction term $M_{\vec{k}}$ is given by:
\begin{eqnarray}
M_{\vec{k}}=\underbrace{g^2S^{2}_3}_{I} -\underbrace{2gk_3S^3 \gamma_5}_{II} +
\underbrace{2gm\Sbar
\gamma_5}_{III} +\underbrace{m^2}_{IV} 
+\underbrace{ig\dot S_3 \gamma^0 \gamma^3 
\gamma_5}_{V}
\end{eqnarray}
From (\ref{bog}), we get for the Bugolubov coefficients to
first order in perturbation theory the following
expression:
\begin{eqnarray}
\beta_{\vec{k}ss^\prime} =  
V^\dagger _{-\vec{k}s^\prime}\psi^{in}_{\vec{k} s}
= -\frac{i}{4\omega^{2}}\int_{-\infty}^{\infty}v^\dagger_{-\vec{k}s^\prime}
M_{\vec{k}} u_{\vec{k}s}e^{-i2\omega t^{\prime}} dt^{\prime}
\label{beta}
\end{eqnarray}
where we have substituted $\psi^{in}_{\vec k s}$ in the r.h.s of
(\ref{pert})  
by the lowest order solution $U_{\vec k s}$ and we have made use of the 
limit $t\rightarrow\infty$ of equation
(\ref{pert}).
The corresponding number of particles in a given spin state is given by the 
sum:  
\begin{eqnarray}
N_{\vec{k}s} = \sum_{s^\prime} |\beta_{\vec{k}s s^{\prime}}|^2
\end{eqnarray}
Now we discuss the contribution of each  term present in the potential.
First we notice that the diagonal terms   $I$ 
and $IV$ do not contribute to first order in perturbation theory.
The  $V$ term  gives rise to:
\begin{eqnarray}
v^\dagger_{-\vec{k}s^\prime}ig\dot S_3\gamma^0\gamma^3\gamma^5 u_{\vec{k} s}
= 2g\dot{S}_3 (k_2
\mp ik_1)
\end{eqnarray}
where $-$ signs applies for $s^\prime = -$ and $s = + $ and the $+$ 
sign for 
opposite spin directions. For $s'=s$ the term vanishes.
The $II$ term gives rise to 
\begin{eqnarray}
-2gv^\dagger_{-\vec{k}s}k_3S_3 \gamma_5 u_{\vec{k} s}
=2gk_3S_3\left(\omega+m-\frac{\omega^2-m^2}{\omega+m}\right)
\end{eqnarray}
Finally the $III$ term yields:
\begin{eqnarray}
2gmv^\dagger_{-\vec{k}s}S_3\gamma^3 \gamma_5 u_{\vec{k} s} 
=-4gmS_3k_3
\end{eqnarray} 
Notice that the contributions from $II$ and $III$ cancel each other and
finally only the term $V$ contributes to the Bogolyubov coefficients. 
Performing the Fourier transform implicit in (\ref{beta}), and using
the spherical coordinates in momentum space $(k,\theta,\phi)$,
we can write the total number
of particles produced with spin $s=\pm$ and  momentum 
$\vec{k}$ as:
\begin{eqnarray}
N_{\vec{k}+} =N_{\vec k -}= g^2\vert \hat S_{3}(\omega)\vert ^2 
\frac{\vec k^2}
{\omega^2} \sin^2 \theta
\end{eqnarray}
with $\hat S^3(\omega)$ the Fourier transform of
$S^3(t)$ and $\theta$ the angle between $\vec k$ and $\vec S$. 
As commented before, the number of produced fermions with spin up
equals the number of fermions with spin down and as a consequence
there is no net spin creation. The total number of fermions is
just twice the above result.
 
We see that the production is not isotropic, with a maximum in the orthogonal
directions to $\vec S$ ($\theta=\pi/2$) and no forward-backward 
production ($\theta=0,\pi$). In Fig.1 we have plotted the spectrum in the
lower curve  
and we observe the resonance close to  $k=m_s/2$ which corresponds
to the decay of the  axial field quanta into  fermion-antifermion pairs.

\newpage
{\bf Massless  fermions}

As we have seen, it is not possible to separate out the positive 
and negative spin states in the massive case, but in the massless case 
it is possible to do that by choosing a different representation for the 
Dirac matrices, the Weyl representation. Since the solutions of the
Weyl equation in the absence of external sources can be written 
as eigenstates of the helicity operator, it is useful to consider those
eigenstates for the positive and negative frequencies solutions.
We shall derive  the total number of
massless fermion production and we shall see that the summation of 
both  chiralities gives rise to total number of fermions and it is equal
to the one obtained in the previous section taking 
$m\rightarrow 0$. The form of the plane-wave solutions with positive and
negative frequency is that given in (\ref{planew}), but now the form of the
spinors is given by:
\begin{eqnarray}
u_{\vec{k}+} =\sqrt{2 \omega}\left( \begin{array}{c}
a_{+}(\vec{k})\\
0\
\end{array}\right),\;\;
u_{\vec{k}-} =\sqrt{2 \omega}\left( \begin{array}{c}
0\\
a_{-}(\vec{k})\
\end{array}\right)
\end{eqnarray}
\begin{eqnarray}
v_{\vec{k}+} =\sqrt{2 \omega}\left( \begin{array}{c}
b_{+}(\vec{k})\\
0\
\end{array}\right),\;\;
v_{\vec{k}-} =\sqrt{2 \omega}\left( \begin{array}{c}
0\\
b_{-}(\vec{k})\
\end{array}\right)
\end{eqnarray}
where 
\begin{eqnarray}
a_{+}(\vec{k})=-b_{+}(\vec{k}) =\left( \begin{array}{c}
\cos\frac{ \theta }{2} \\
\cos \frac{\theta}{2} e^{i\phi }\
\end{array}\right ),\;\;
a_{-}(\vec{k})=-b_{-}(\vec{k}) =\left( \begin{array}{c}
-\sin\frac{\theta}{2} e^{-i\phi }\\
\cos\frac{ \theta }{2} \
\end{array}\right )
\end{eqnarray}
are eigenstates of the helicity operator with eigenvalues $+1$ and
$-1$ respectively.
The normalization conditions  are the same as the massive ones.
In the Weyl representation the gamma matrices are:
\begin{eqnarray}
\gamma^{0}=\left( \begin{array}{cc}
0 & -1\\
-1&0\
\end{array}\right ),\;\;
\gamma^{i}=\left( \begin{array}{cc}
0 & \sigma^{i}\\
-\sigma^{i}&0\
\end{array}\right ),\;\;
\gamma_{5}=\left( \begin{array}{cc}
1 & 0\\
0& -1\
\end{array}\right )
\end{eqnarray}
In the present case, the initial positive frequency solution will evolve into
a linear combination of positive and negative frequency modes, but 
since the perturbation is diagonal in chirality space there is
no mixing between positive and negative chirality modes, i.e:
\begin{eqnarray}
\psi^{in}_{\vec{k} s}(t,\vec{x})\stackrel{t\rightarrow \infty}{\rightarrow}
\alpha_{\vec{k}s}
U_{\vec{k}s} + \beta_{\vec{k}s}V_{-\vec{k}s} 
\end{eqnarray}
In the massless case, the contributions in the $III$ and
$IV$ terms are not present and again the diagonal terms do not
contribute. As a consequence only the $V$ term is 
relevant. Hence the total number of 
fermions with momentum $\vec k$ and helicity $s$ is given by the 
following expression:
\begin{eqnarray}
N_{\vec{k} s}= \vert \beta_{\vec k,s}\vert ^2
=g^2\sin^{2} \theta  \vert \hat{S_3}(\omega)\vert ^2
\end{eqnarray}
The total number with both the spins would be just  twice the above
expression, which is the same as the massive case. In Fig. 1 we have 
represented the
spectrum for massless fields in the upper curve.
It is to be noted that the 
production of massive fermions is, as expected, suppressed by the mass term.
We also observe the peak at $k=m_s/2$.

\begin{figure}
\begin{center}
\mbox{\epsfysize=10cm\epsfxsize=10cm
\epsffile{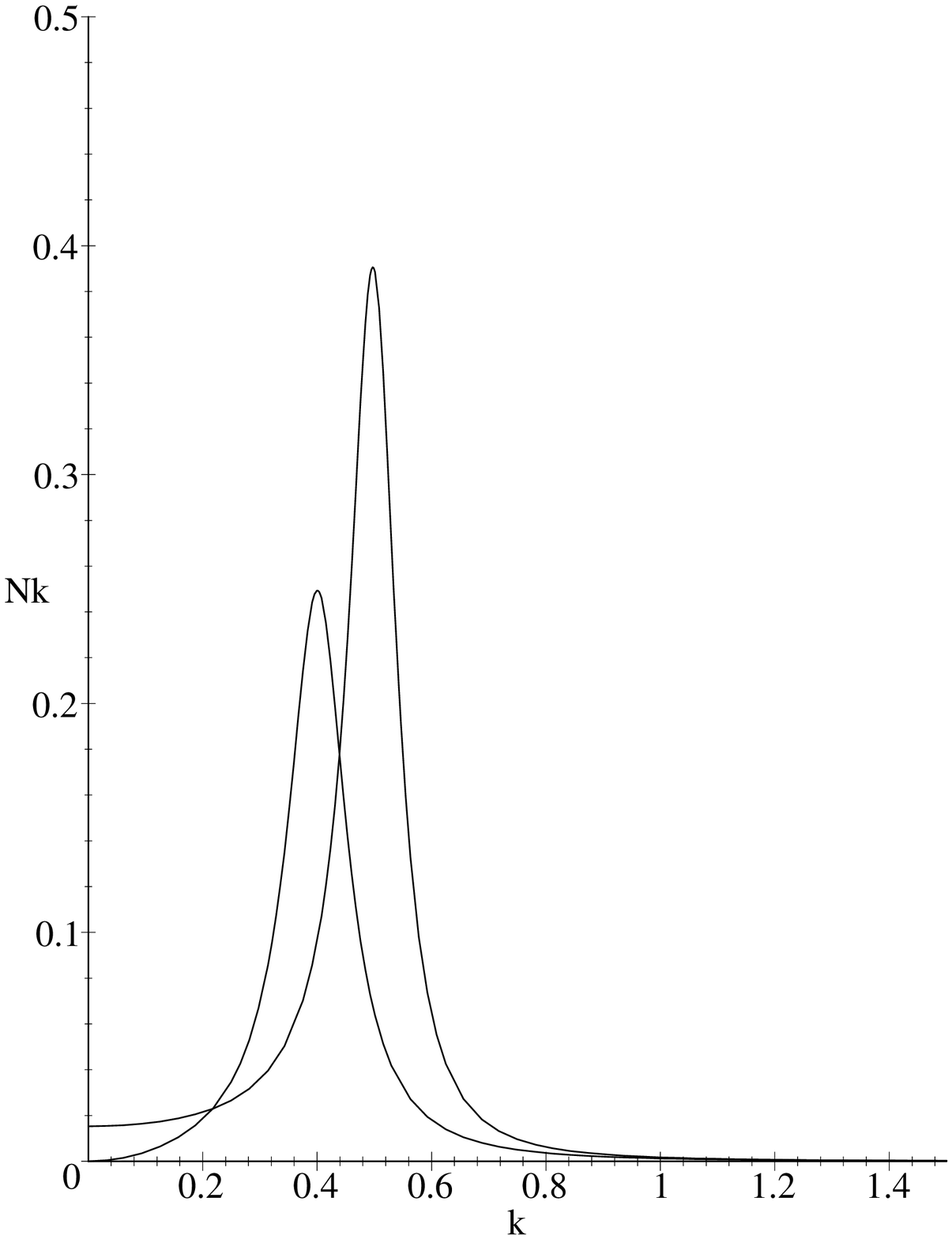}}

\end{center}
\vspace {-.8cm}
\leftskip 1cm
\rightskip 1cm
{\footnotesize
{\bf Figure 1.-}Number of particles against $\vert\vec k\vert$ 
for $g\hat S=1/8\; m_s$ and $\Gamma=0.2\; m_s$. For massive 
fermions $m=0.3m_s$ } 
\end{figure}

\section{Non-perturbative results}

In the previous section we have obtained the spectrum and
angular distribution of the produced particles up to
first order in perturbation theory. Higher order terms
are expected to be suppressed  and
only will give rise to smaller peaks at higher energies. 
However, the non-perturbative effects can have a more important
effect on the results. As shown in the scalar case, the features
of production essentially deviate from the pertubative result. 
Since in our case the equations cannot be reduced to any known form, 
we will solve them numerically. For the sake of simplicity
we will only consider the massless case in which it is possible to
disentangle the two spin states. We will also assume that the 
behaviour of the background field is the following: for $t<0$ and
$t>nT$ 
$S_3=\hat S$ a constant, with $n$ an integer and $T=2\pi/m_s$ the
period of the oscillations. For $0<t<nT$ $S_3=\hat S \cos(m_st)$. 
It has been checked that the final 
spectra do not depend on $n$.
In order to reduce (\ref{direq}) to a second order form we make
the following ansatz on the solutions
\cite{GK}:
\begin{eqnarray}
\psi=(i\pabar+g\Sbar\gamma_5)f_{\vec k s}(t)e^{i\vec k \vec x}W_{s \pm} 
\end{eqnarray}
where for positive(negative)+(-) energy modes with spin s we take:
\begin{eqnarray}
W_{s+}=\left(\begin{array}{c}
\chi_s\\0\end{array}\right),\;\;
W_{s-}=\left(\begin{array}{c}
0\\\chi_s\end{array}\right)
\end{eqnarray}
In this section we will work again in the Weyl representation.
With this ansatz, the equations of motion can be written as a
single equation for the function $f_{\vec k s}$.
\begin{eqnarray}
(\partial_0^2+\vec k^2-isg\dot S_3+2gk_3S_3+g^2S_3^2)f_{\vec k s}(t)=0
\label{second}
\end{eqnarray}
In addition the $\psi$ spinors will give rise 
asympotically ($t\rightarrow \pm \infty$) to eigenstates of the
the third component of the spin operator. In this sense, they have
the appropriate form to describe the creation of fermions with
definite spin along that direction.
The normalized 
spectrum of the total number of particles created with spin $s$ is
given by (see \cite{moste,rusos} for details):
\begin{eqnarray}
N_{\vec k s}&=&\frac{k_1^2+k_2^2}{\vec k^2+g^2\hat S^2+2g\hat S k_3}
\frac{1}{\sin^2(d_2)}
(\im F_{\vec k s}(T))^2\nonumber \\
&=&\frac{k^2\sin^2(\theta)}{k^2+g^2\hat S^2-2g\hat S k
\cos \theta}
\frac{(\im F_{\vec k s}(T))^2}{1-(\re F_{\vec ks}
(T))^2}
\end{eqnarray}

\begin{figure}
\vspace{-1cm}
a)\mbox{\epsfysize=6cm\epsfxsize=6cm
\epsffile{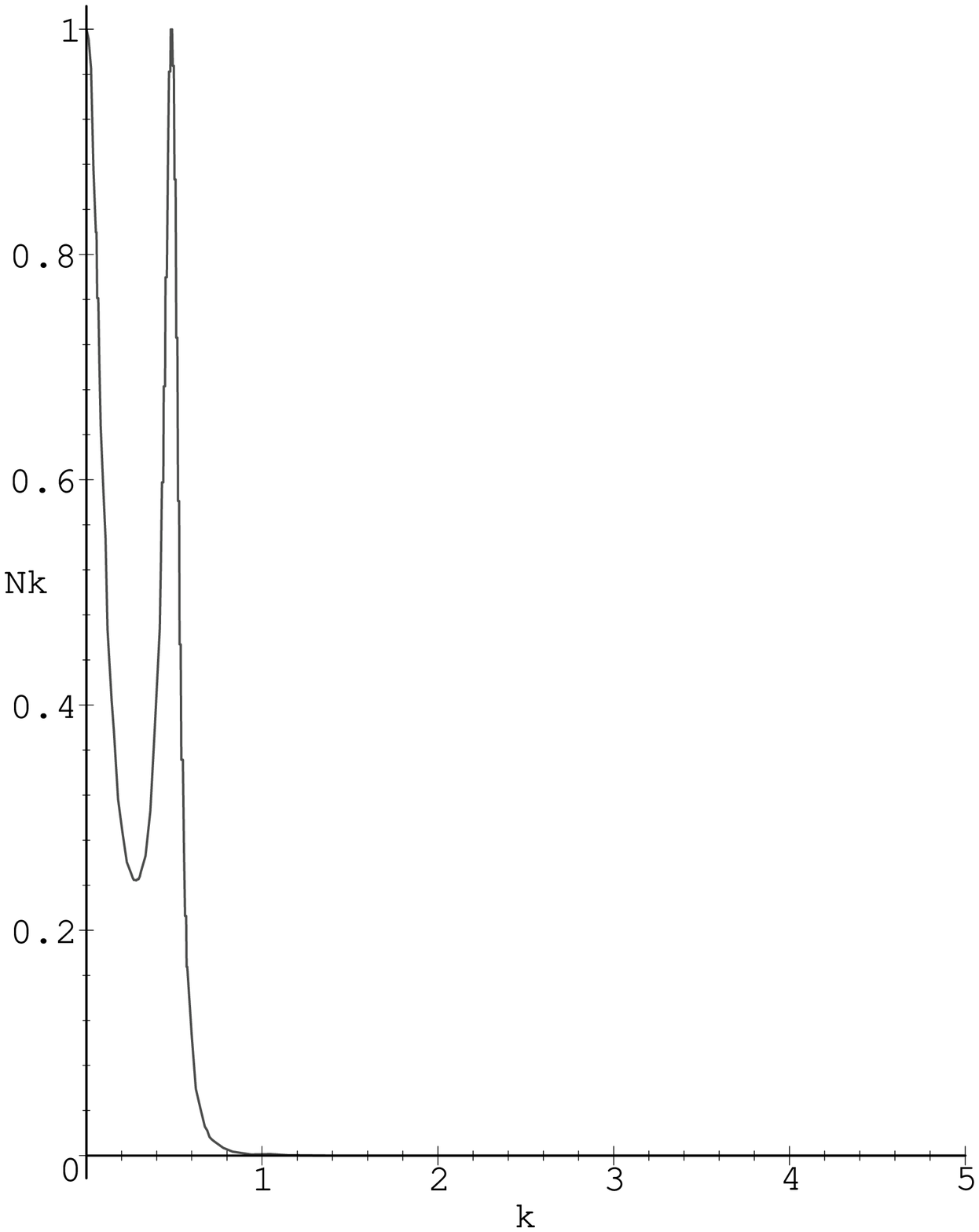}}
b)\mbox{\epsfysize=6cm\epsfxsize=6cm
\epsffile{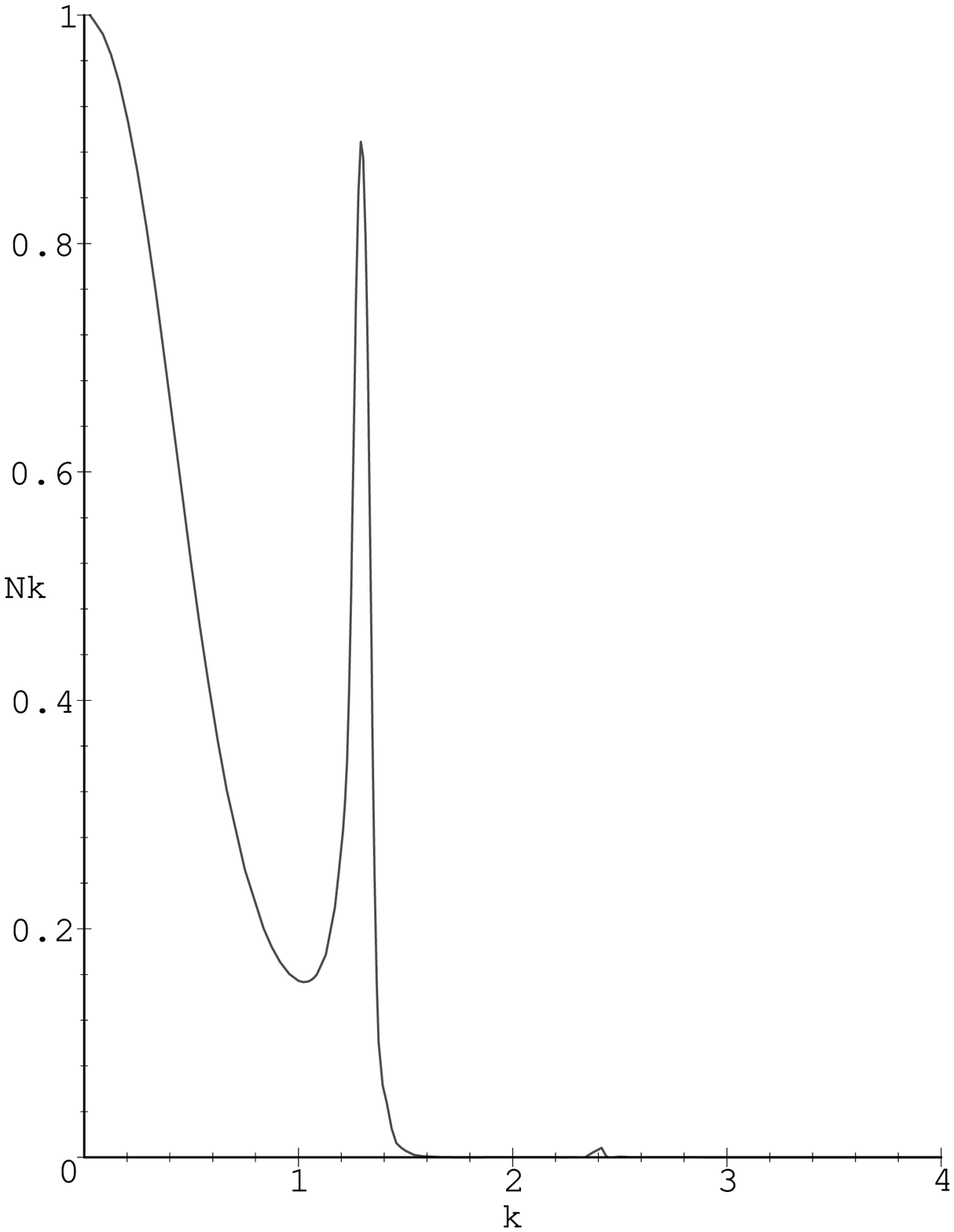}}\\
c)\mbox{\epsfysize=6cm\epsfxsize=6cm
\epsffile{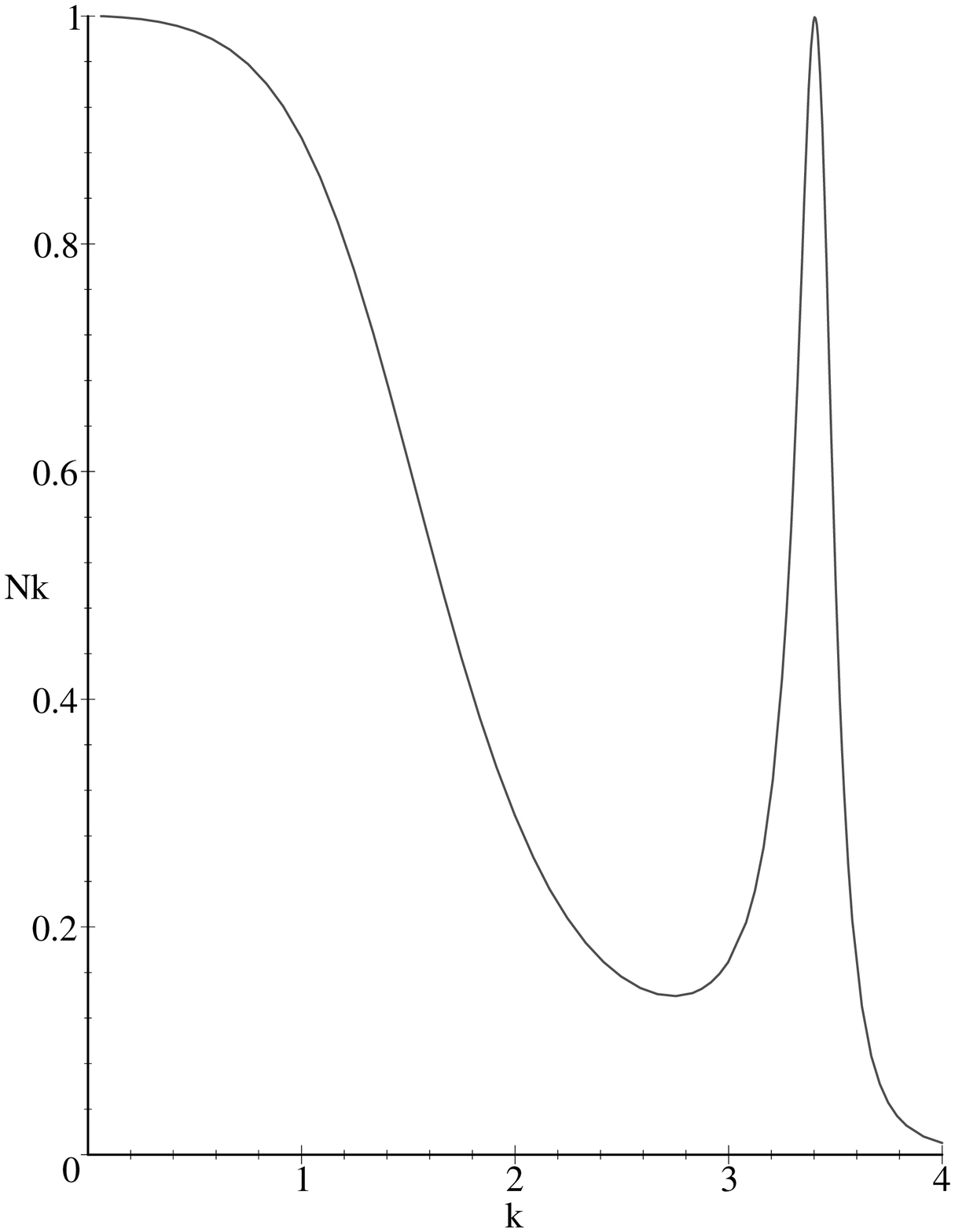}}
d)\mbox{\epsfysize=6cm\epsfxsize=6cm
\epsffile{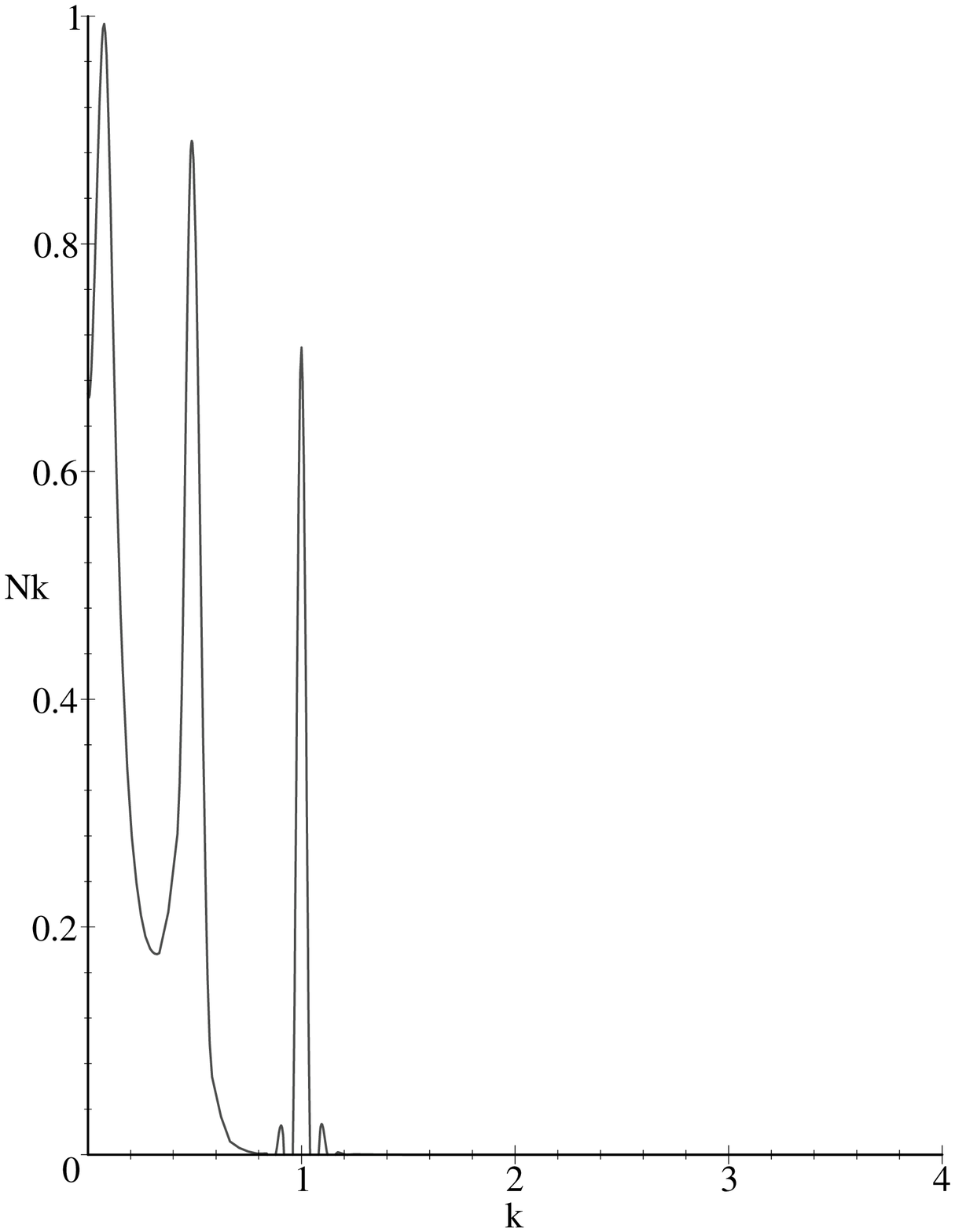}}\\
e)\mbox{\epsfysize=6cm\epsfxsize=6cm
\epsffile{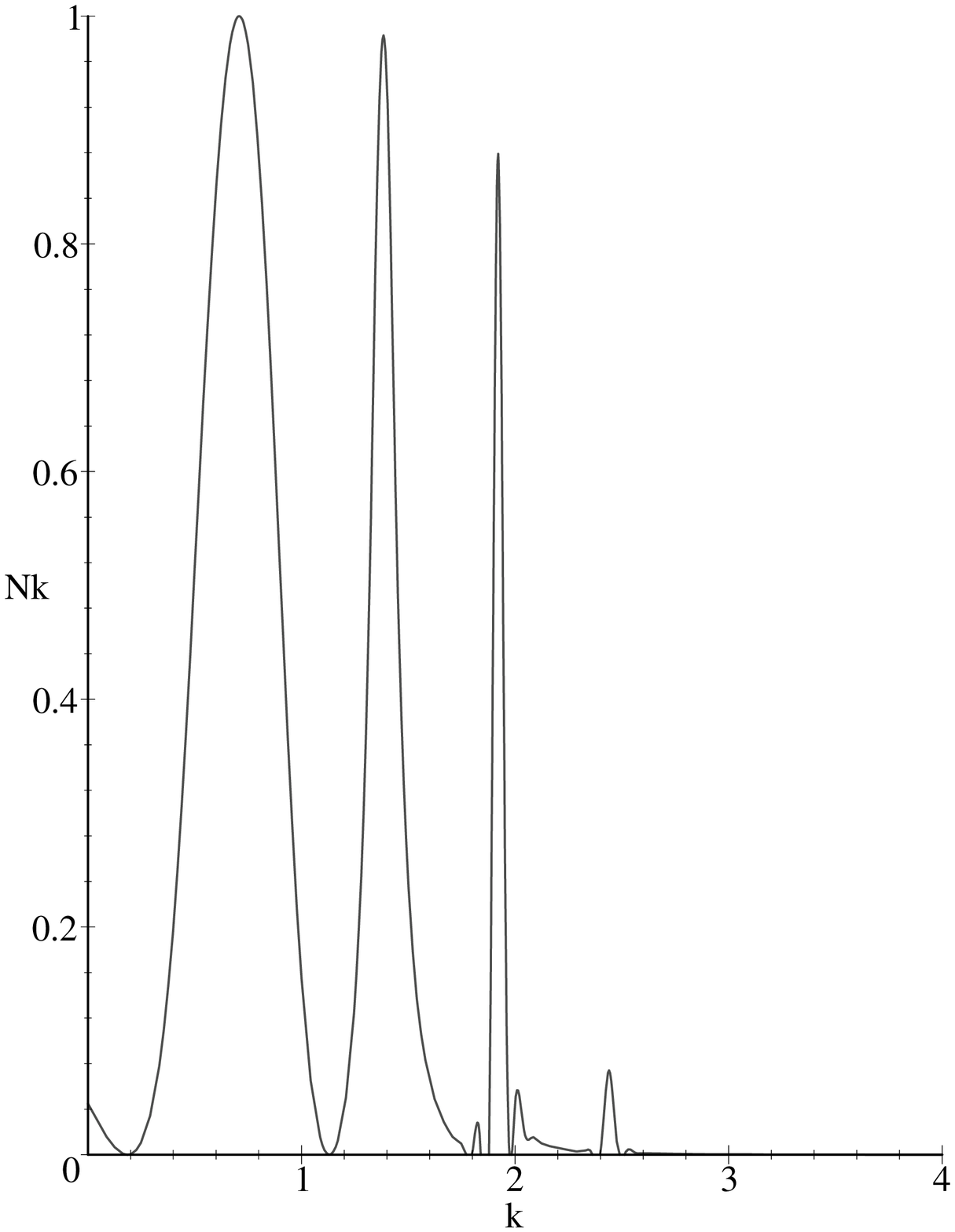}}
f)\mbox{\epsfysize=6cm\epsfxsize=6cm
\epsffile{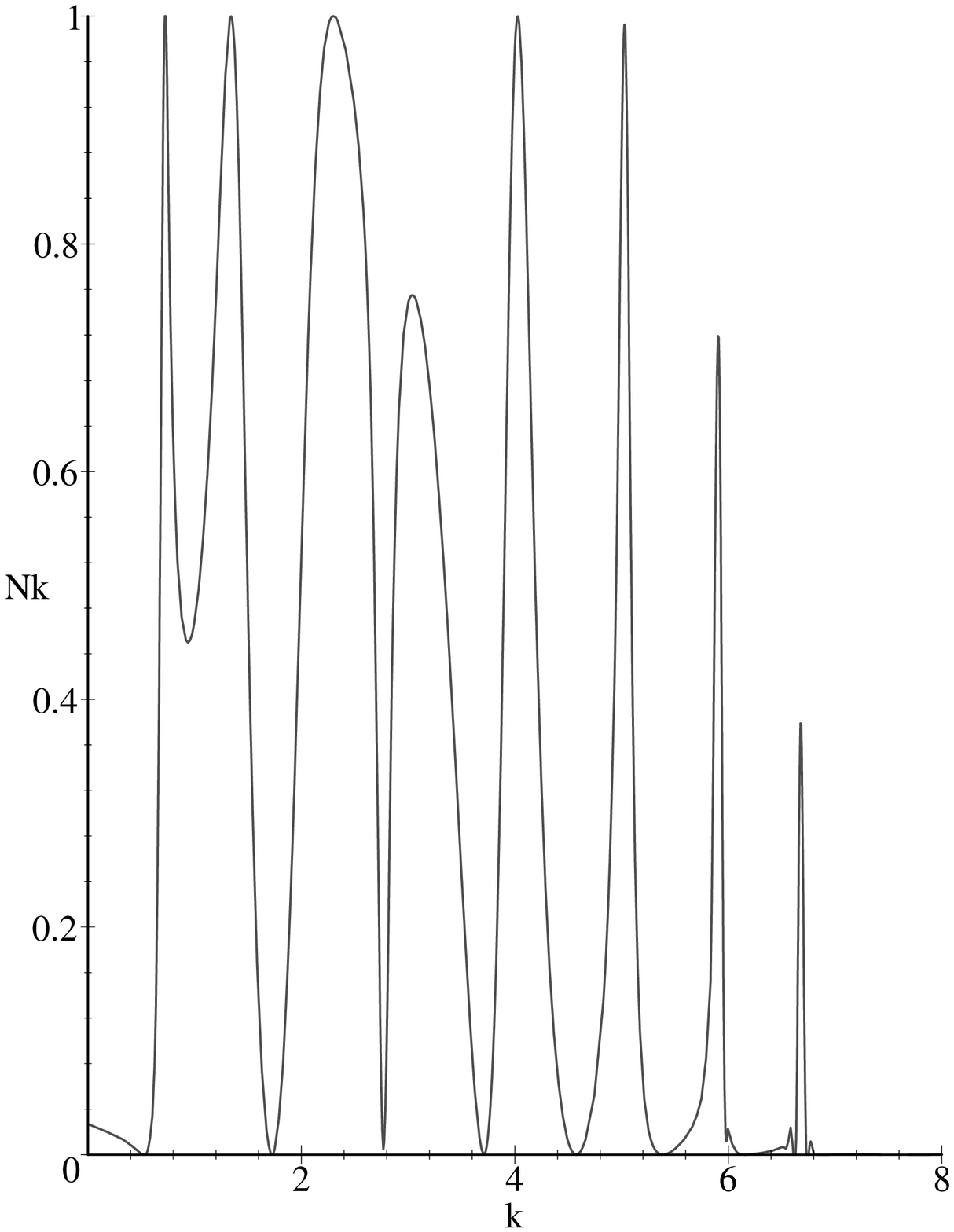}}
\leftskip 1cm
\rightskip 1cm
{\footnotesize
{\bf Figure 2.-}Normalized number of particles against $\vert \vec k \vert$.
a),b) and c) correspond to $g\hat S=0.1, 1 ,10m_s$ 
 respectively with $\theta=\pi/2$
and d), e) and f) to the same amplitudes but for $\theta=\pi/4$.} 
\end{figure}

\begin{figure}
\vspace{-1.3cm}
a)\mbox{\epsfysize=6cm\epsfxsize=6cm
\epsffile{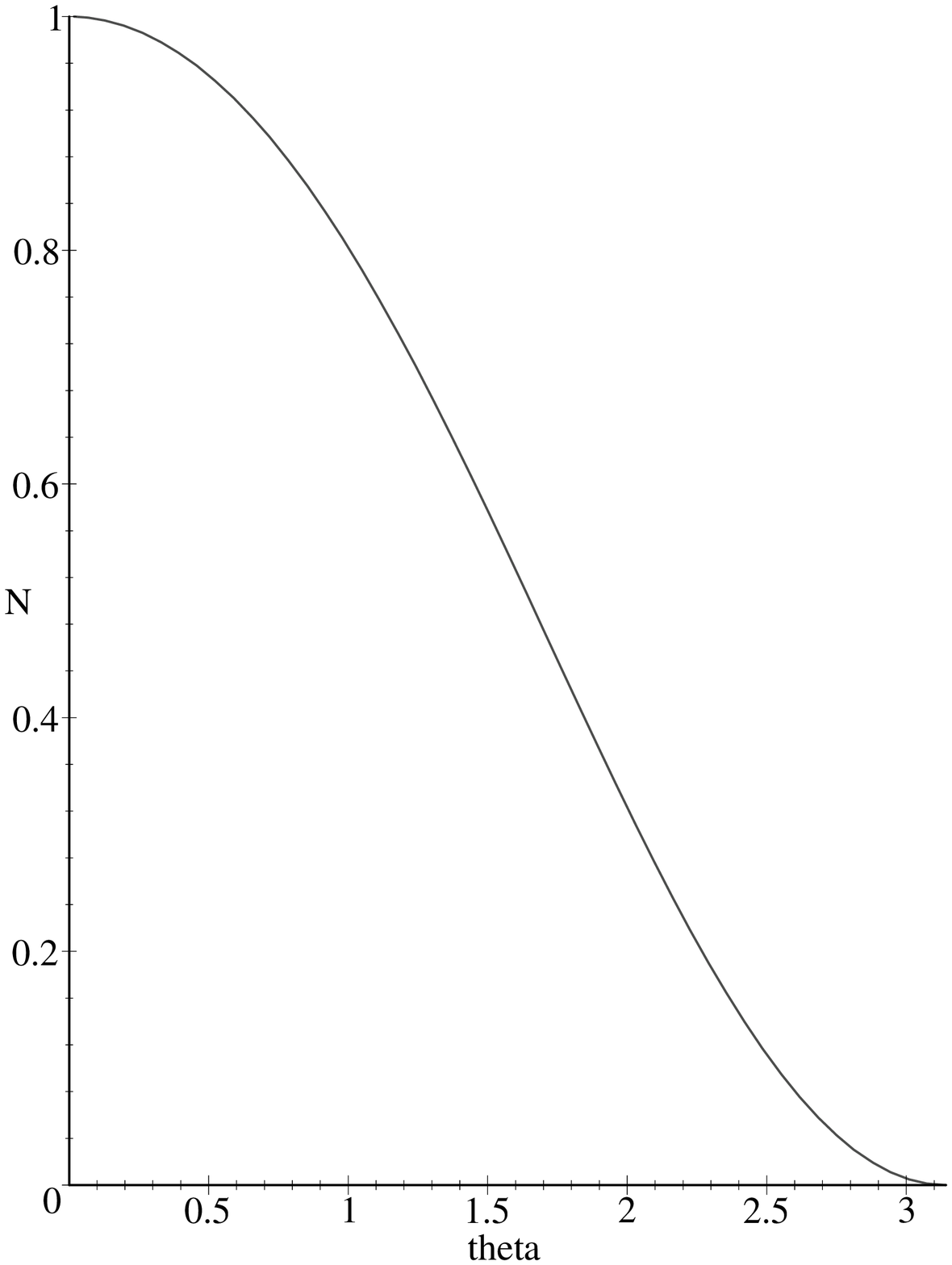}}
b)\mbox{\epsfysize=6cm\epsfxsize=6cm
\epsffile{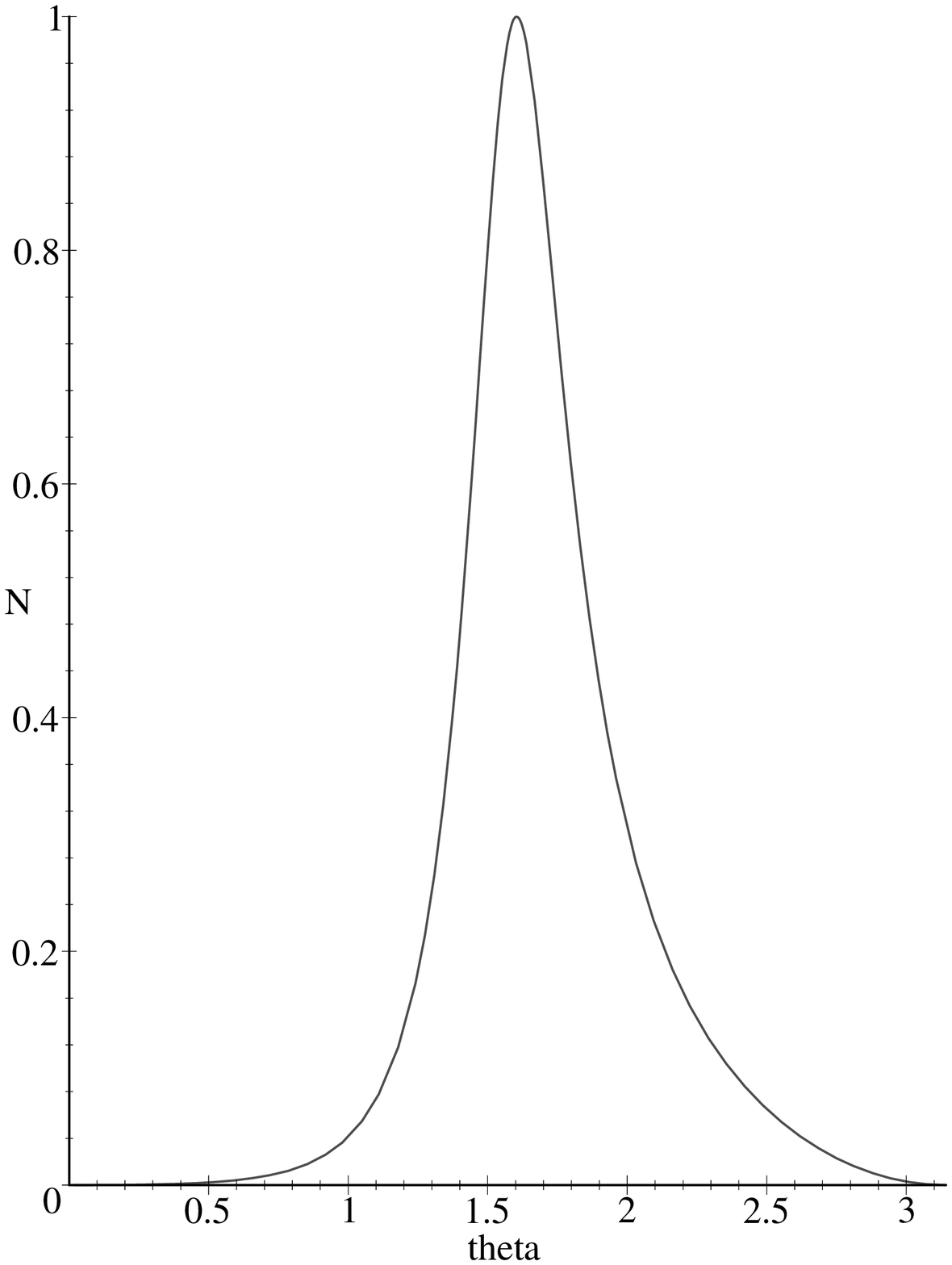}}\\
c)\mbox{\epsfysize=6cm\epsfxsize=6cm
\epsffile{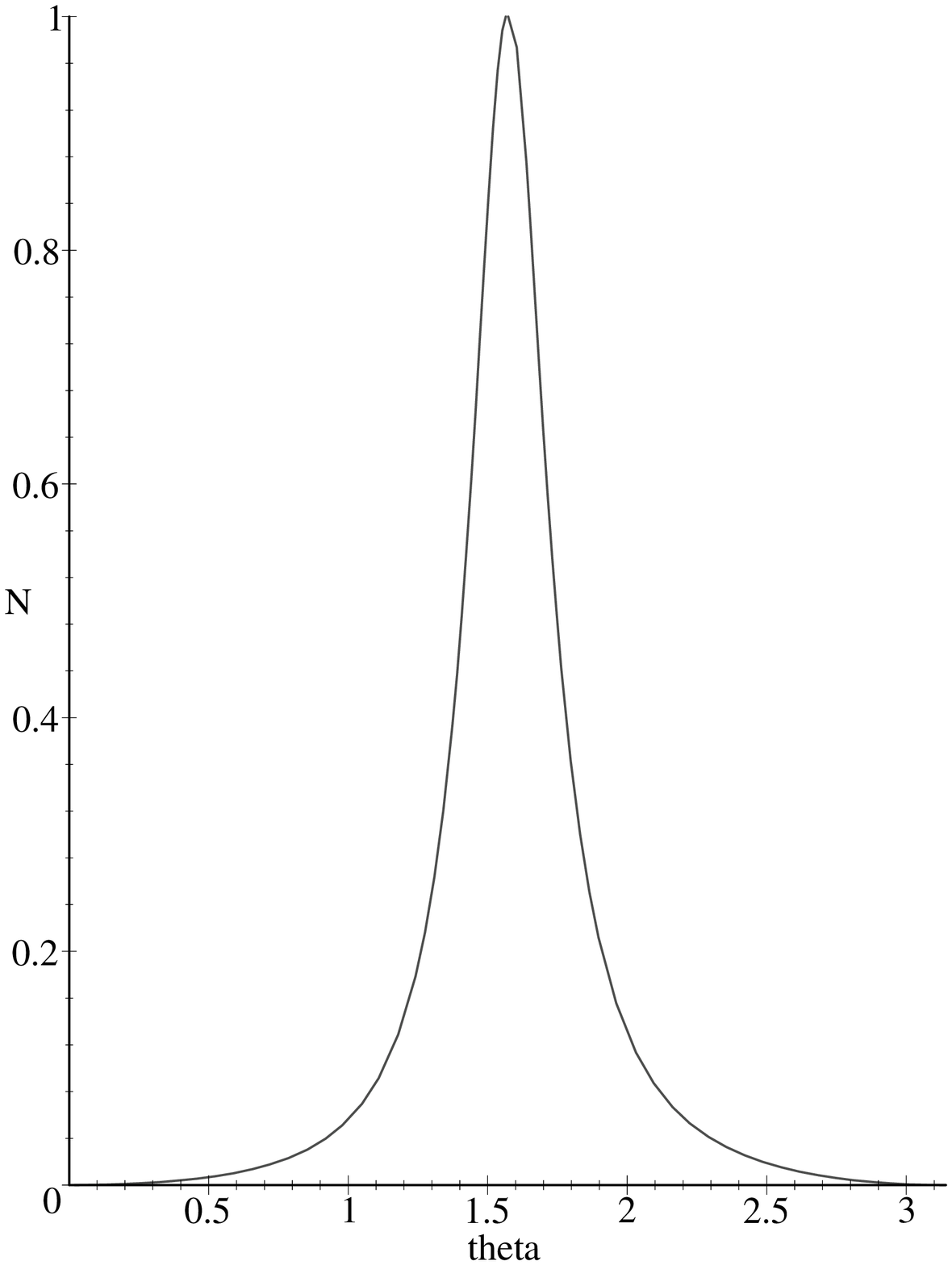}}
d)\mbox{\epsfysize=6cm\epsfxsize=6cm
\epsffile{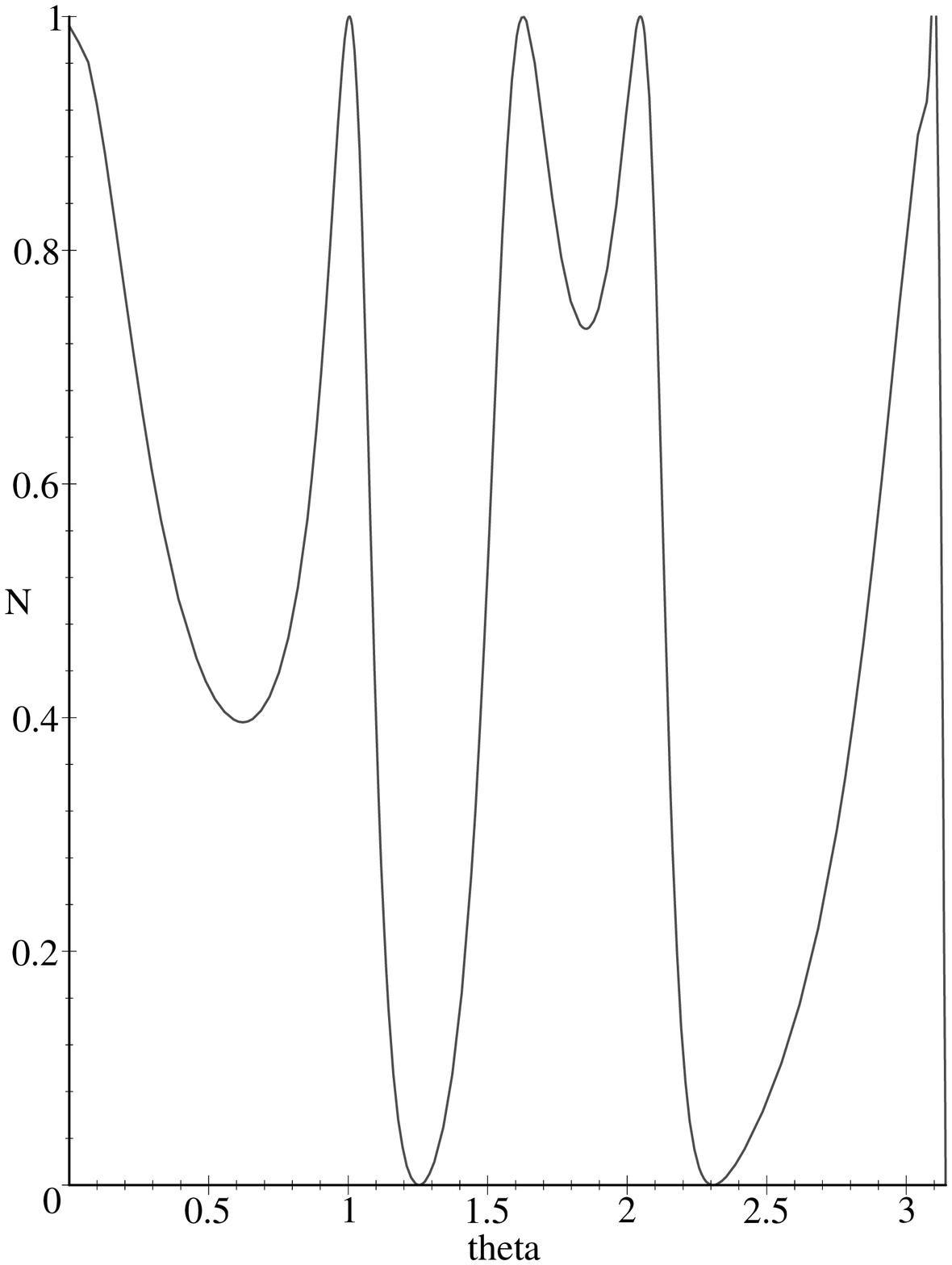}}\\
e)\mbox{\epsfysize=6cm\epsfxsize=6cm
\epsffile{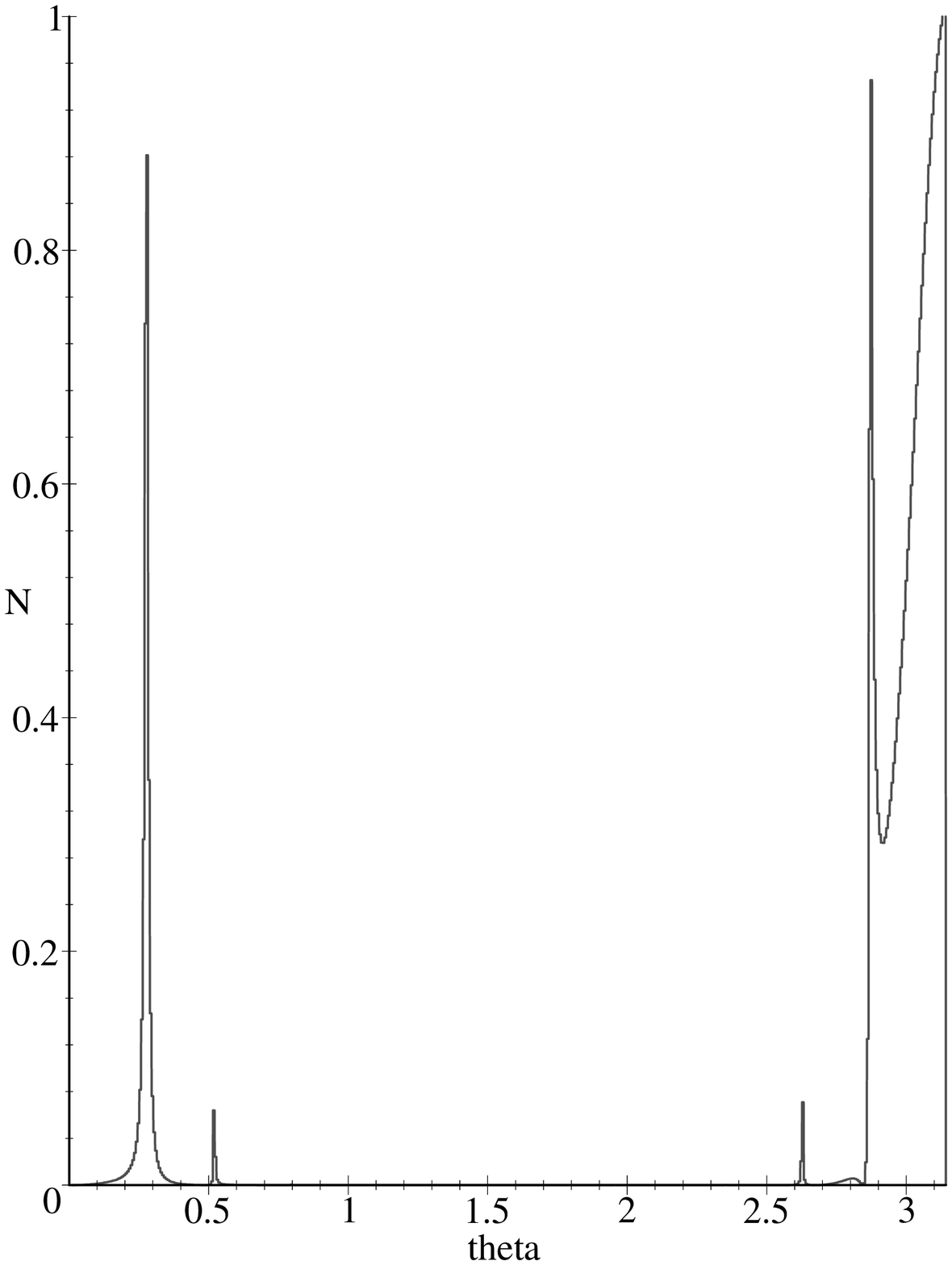}}
\leftskip 1cm
\rightskip 1cm

{\footnotesize
{\bf Figure 3.-}Normalized number of particles against $\theta$.
a),b) and c) correspond to $g\hat S=0.1, 1 ,10m_s$ 
respectively with 
$\vert \vec k \vert=.1m_s$
and d), e)  to $g\hat S=10m_s$ with $\vert \vec k \vert=1, 10m_s$ 
respectively.} 
\end{figure}

\noindent where $F_{\vec k s}$ is a solution of equation (\ref{second}) 
with initial conditions $F_{\vec k s}(0)=1$ and $\dot F_{\vec k s}(0)=0$ and 
we have used the fact that $\cos d_2=\re F_{\vec k s}(T)$.
The numerical computations show that the number of fermions produced
with spin up is the same as those with spin down, as expected
there is no net spin creation in this particularly simple model.
We see that $\sin^2 \theta$ angular dependence that we obtained in
the perturbative calculation is already present in the above result
for $\hat S \gg k$.
Although it is expected to be modified by the factor involving 
$F_{\vec k s}(T)$. In Figures 2 and 3 we have represented the spectra 
against $\vert \vec k \vert$ and $\theta$  
for different values of the amplitudes.
First we notice that for small amplitudes of the oscillations
$\hat S$, there appears a peak at
$k=0.5 m_s$ in agreement with the perturbative calculation.
However we find new peaks also for small amplitudes. For large 
values of the amplitude the departure from the perturbative 
results is apparent.
We can also observe the appearance of the resonance bands in the
spectra whose structure  strongly depends on the angle $\theta$. 
In addition, the angular dependence is also highly sensitive to the 
energy $k$ of the fermions and the amplitude $\hat S$ of the 
oscillations. In particular we find that for large amplitude of
the oscillations,  strong peaks appear  at some particular
values of the angles, indicating that the production mainly takes place
in those directions.

\section{Conclusions}

In this work we have studied the resonant production of fermions
from a classical oscillating axial background. The dynamics of the
classical background is determined by the renormalizabililty of the
quantum theory of fermions coupled to axial fields and is given
by the Proca equation. The oscillatory solutions to this equation
are obtained and the corresponding phenomenon of resonant particle 
production is analyzed both from a perturbative and a non-perturbative
point of view. In the perturbative case, we have considered both
the massive and massless fermion cases for a simple background in
which the axial field is oscillating in the z-direction, and we have 
obtained the spectra and angular distributions of the produced particles.
In the non-perturbative case, we consider only the massless case and
study the equation numerically. We have checked that the non-perturbative
results deviate from the perturbative ones with the appearance
of the typical resonance bands, that in our case depend both
on momenta and angle.

Concerning the possibility that this mechanism could have been 
relevant in the early universe, we should mention that the generation
of an initial anisotropic axial field that later on could have evolved 
according to (\ref{proca}) has been explored in \cite{andrianov}. 
There it is shown that for axial fields interacting with photons, there is
a possibility of  
spontaneous symmetry breaking along spatial directions for the axial field.
As a consequence, that field  would acquire some spatial-like vacuum 
expectation value.
In addition, as we mentioned in the introduction, axial fields are  
present in the context of string
cosmology and therefore it would be very interesting to study  
if the resonant phenomenon studied in this work could affect  
the standard picture of reheating after inflation.

\vspace{0.5cm}
{\bf Acknowledgments:} A.L.M. acknowledges support from  SEUID-Royal Society
and (CICYT-AEN96-1634)(Spain). A.M. is supported by INLAKS scholarship 
and an ORS award. We thank Juan Garc\'{\i}a Bellido, Luis Mendes 
and Andrew Liddle for valuable
discussions.


\end{document}